# Derivation of the Planck Units Based in a Membranes Model

**Lizandro B. R. Zegarra[1] and Luis E. G. Armas[2]**

[1]Departamento de Matemática de la Universidad Nacional del Santa – UNS, 508, Chimbote – Perú

[2]Grupo de Óptica Micro e Nanofabricação de Dispositivos (GOMNDI), Universidade Federal do Pampa, Campus Alegrete, 97546-550, Rio Grande do Sul - RS, Brazil

*Corresponding Authors*: Lizandro B. R. Zegarra
[1] e-mail: lreynazegarra@gmail.com

Luis E. G. Armas
[2] e-mail: luisarmas@unipampa.edu.br

## Abstract

In this study, the Planck units (mass, time and length) have only been derived, explained and attributed a physical meaning when they were deduced based on the concept of interacting membranes (membranes instead of strings of string theory). For this purpose, a set of five assumptions were proposed: (a) the existence of the interacting membranes; (b) the curvatures of the membranes oscillate according to the classical wave equation; (c) the spatial period of the wave that arise when the membranes oscillate is given by $\lambda = \xi\pi / k$; (d) the membranes oscillate with wavelength given by de Broglie relation and (e) $x = ct$ holds**.** The parameter $\xi$ determines the period of oscillation of the given membranes. In deriving the Planck units in this work, $\xi$ must take the value 2 and determines a period $2\pi$, closely to minimum value 1 or to fundamental period $\pi$, respectively. In this context, Planck units must be fundamental. Moreover, the parameter $\xi$ was reported as a unification parameter between the formulas for the Coulomb's law and Newton's law of universal gravitation linking the forces of microworld and macroworld. Depending on the value $\xi$ takes, one force or another will be had. It is also shown that the potential $V = hc / \xi\pi x$ deduced from the above assumptions and which contributes to deduce the Planck units, can be derived from Yukawa's equation. Hence, the present work would be contributing to theoretical physics, since at the Planck scale predictions of some theories like Standard Model, quantum field theory and general relativity are not expected to be valid.

**Résumé**

Dans cette étude, les unités de Planck (masse, temps et longueur) n'ont été dérivées, expliquées et dotées d'une signification physique que lorsqu'elles ont été déduites sur la base du concept de membranes en interaction (membranes au lieu de cordes dans la théorie des cordes). À cette fin, un ensemble de cinq hypothèses a été proposé : (a) l'existence de membranes en interaction ; (b) les courbures des membranes oscillent selon l'équation classique des ondes ; (c) la période spatiale de l'onde qui apparaît lorsque les membranes oscillent est donnée par $\lambda = \xi\pi / k$ ; (d) les membranes oscillent avec une longueur d'onde donnée par la relation de De Broglie et (e) $x = ct$ est satisfait. Le paramètre $\xi$ détermine la période d'oscillation des membranes données. Dans ce cas $\xi$ prend la valeur 2 et détermine une période $2\pi$, proche respectivement de la valeur minimale 1 ou de la période fondamentale $\pi$. Dans ce contexte, les unités de Planck doivent être fondamentales. De plus, le paramètre $\xi$ a été signalé comme un paramètre unificateur entre les formules de la loi de Coulomb et de la loi de gravitation universelle de Newton reliant les forces du micromonde et du macromonde. Selon la valeur que prend $\xi$, on obtiendra une force ou une autre. On montre également que le potentiel $V = hc / \xi\pi x$ déduit des hypothèses précédentes et qui contribue à dériver les unités de Planck, peut être dérivé de l'équation de Yukawa. Par conséquent, le présent travail contribuerait à la physique théorique, puisqu'à l’échelle de Planck, les prédictions de certaines théories telles que le modèle standard, la théorie quantique des champs et la relativité générale ne devraient pas être valides.

## 1. Introduction

The Planck units, also known as natural units, were proposed in the year 1899 by Max Planck, so that the well – known fundamental constants of physics among others such as: speed of light $(c)$, reduced Planck constant $\left(\hbar = \frac{h}{2\pi}\right)$ and universal gravitational constant $(G)$ take the value of 1 when they are present in the calculations.

Max Planck defined a natural scale from dimensional analysis for time, length and mass composed of physical constants involving only the speed of light $c$, the gravitational constant $G$ and the (later defined as) reduced Planck constant $\hbar$ given by[1].

$$t_P = \sqrt{\frac{\hbar G}{c^5}} \ , \ l_P = \sqrt{\frac{\hbar G}{c^3}} \ , \ m_P = \sqrt{\frac{\hbar c}{G}}$$

Weinstein and Rickles[1] state that the Planck units are believed to be crucial for a quantum theory of gravity, but apart from a dimensional analysis, a deeper understanding of this link from these units is missing and a successful quantum gravity theory is still to be developed. Meschini[2] states: Dimensional analysis is a surprisingly powerful method capable of providing great insight into physical situations without needing to work out or know the detailed principles underlying the problem in question. This (apparently) suits research on quantum gravity extremely well at present, since the physical mechanisms to be involved in such a theory are unknown. However, dimensional analysis is not an all-powerful discipline: unless very judiciously used, the results it produces are not necessarily meaningful and, therefore, they should be interpreted with caution.

Crothers and Dunning-Davies[3] argue about Planck units: It seems to have been used initially as a means of making equations and expressions dimensionless by making use of suitable combinations of the universal constants $c$, the speed of light, $G$, Newton's universal constant of gravitation, and finally Planck's constant.

Wang et al.[4] state regarding Planck units: They are given corresponding physical meanings, such as: the mass of the ground state particles cannot be greater than the Planck mass; the energy cannot be greater than or equal to the Planck energy, otherwise it will collapse into a black hole; Planck time is an observable event Minimum process time; Planck length is a measure of Planck's quality black hole, unable to distinguish events within a distance less than Planck length; unable to describe events occurring within Planck time when the universe was born, etc. However, the Planck Unit cannot resolve the logical contradiction with the continuous space – time of the theory of relativity, because the space – time structure derived from Planck's length and Planck's time is discontinuous. Even for quantum field theory, Planck length and point particle models are in conflict.

Perl[5] states: $M_P$ is sometimes given as the maximum mass at which a particle can be defined because the gravitational force of the Planck particle distorts too much the space in which we define an elementary particle. I am not convinced that $M_P$ has anything to do with the upper limit on elementary particle masses. $M_P$ mixes a constant from quantum mechanics with two classical constants. So, it indeed marks some sort of boundary between classical and quantum physics and may have to do with the problem of the unification of gravitational and electromagnetic forces. But I am not convinced that $M_P$ has anything to do with elementary particle masses.

Buczyna[6] state about Planck units: The original derivation of the Planck "base" units of mass, length and time was via dimensional analysis. Attributing physical basis for the existence of such quantities was motivated by the post-quantum field theory expectation of a quantum gravity theory. Apart from defining fundamental scales of length, mass, and time due to fundamental quantum gravitational constraints on space-time and matter, such an encompassing theory can be expected to provide a basis for microscopic physical description of all of physics.

Pankovic[7] has determined the basic Planck units (mass, length, and charge) using two dynamical principles. The first principle corresponds to the definition of the Compton length of the physical system and the second principle to the definition of the classical dimension of a physical system through the equivalence of the rest energy of a system and the potential energy for the corresponding length. It also shows analogous results when changing the second dynamic principle for new dynamic principles in which the Planck length corresponds to the geometric mean value between the Compton wavelength and the Schwarzschild radius for an arbitrary physical system. And more generally, it shows that the basic Planck units mentioned above can be obtained by analogous relations between three characteristic length parameters determined by general relativity dynamics in the case of the Kerr-Newman black hole with angular momentum equivalent to reduced Planck constant.

According to the results reported on the literature, up to now, it was not attributed a physical meaning to the Planck units, since they were initially deduced from a purely dimensional analysis. Therefore, in the present research work, the Planck units were deduced based on the concept of interacting membranes (membranes instead of strings of string theory), explaining and attributing a physical meaning to them. A parameter $\xi$ which plays a fundamental role in the derivation of the Planck's units was reported in.[8] This parameter unifies the formulas for the Coulomb's law and Newton's law of universal gravitation, linking the forces of microworld and macroworld. Depending on the value $\xi$ takes, one force or another will be had. The universality of the Planck's units which is seriously questioned in[9] is not addressed in this article and should be treated as a lot of subtlety in another article. Here, the presence of the Planck constant in the Planck's units has the same status as it is the case in the de Broglie relation $\lambda = \frac{h}{p}$, or as in $E = h\nu$ for an energy photon in the photoelectric effect.

In the context of this work, we head to contribute to solve some difficulties in the different branches of theoretical physics, since at the Planck scale predictions of some theories like Standard Model, quantum field theory and general relativity are not expected to be valid. Furthermore, one could be faced with the possibility of giving explanations to other unsuspected physical phenomena.

## 2. Theoretical development

Zegarra et al.[8] have shown a simple unification of the classical formulas for the forces of Coulomb's law of electrostatic interaction and Newton's law of gravitational interaction, unification based on the following four assumptions:

First assumption**:** there are two classes of fundamental entities that interact with each other. The first is a sphere (S) with positive gaussian curvature and the second is the surface of a torus (T) there where its gaussian curvature is negative. These entities are called membranes.

Second assumption**:** the curvatures $\left(K\right)$ of the sphere and of the torus oscillate according to the classical wave equation given by

$$\frac{\partial^2 K}{\partial t^2}=v^2\frac{\partial^2 K}{\partial x^2}$$

and with solution given by

$$K=K_0\sin^2\left(kx-\omega t\right)$$

where $v$ is the wave propagation velocity, $k$ is a wave number and $\omega$ the angular velocity.

Third assumption**:** the spatial period or wavelength $\lambda$ of the travelling wave with velocity $v$, originated due to the disturbance of the medium (filled with membranes) that propagates through it when the curvature of the spheres oscillates, is given by $\lambda=\frac{\xi\pi}{k}$. $\xi$ is a dimensionless parameter and $k$ a wave number, as shown in Fig. 1.

Fourth assumption**:** the wavelength with which the curvature of the sphere after a compactification process (when $a\to 0$, $\omega_S=\omega_T$ and $k_S-k_T\to 0$) oscillates, is given by the de Broglie relation $\lambda=\frac{h}{p}$ , as shown in Fig. 1.

Figure 1 shows oscillating spheres with wavelength $\lambda=\frac{h}{p}$ such that $\lambda=\frac{\xi\pi}{k}$ is the wavelength of the originated travelling wave. These wavelengths depend on each point $\left(x,t\right)$. At first, the wavelengths of each oscillating sphere in each point $\left(x,t\right)$ are different.

But, when the wavelengths $\lambda = \frac{h}{p}$ of the all-oscillating spheres are the same, we have a travelling wave with wavelength $\lambda = \frac{\xi\pi}{k}$.

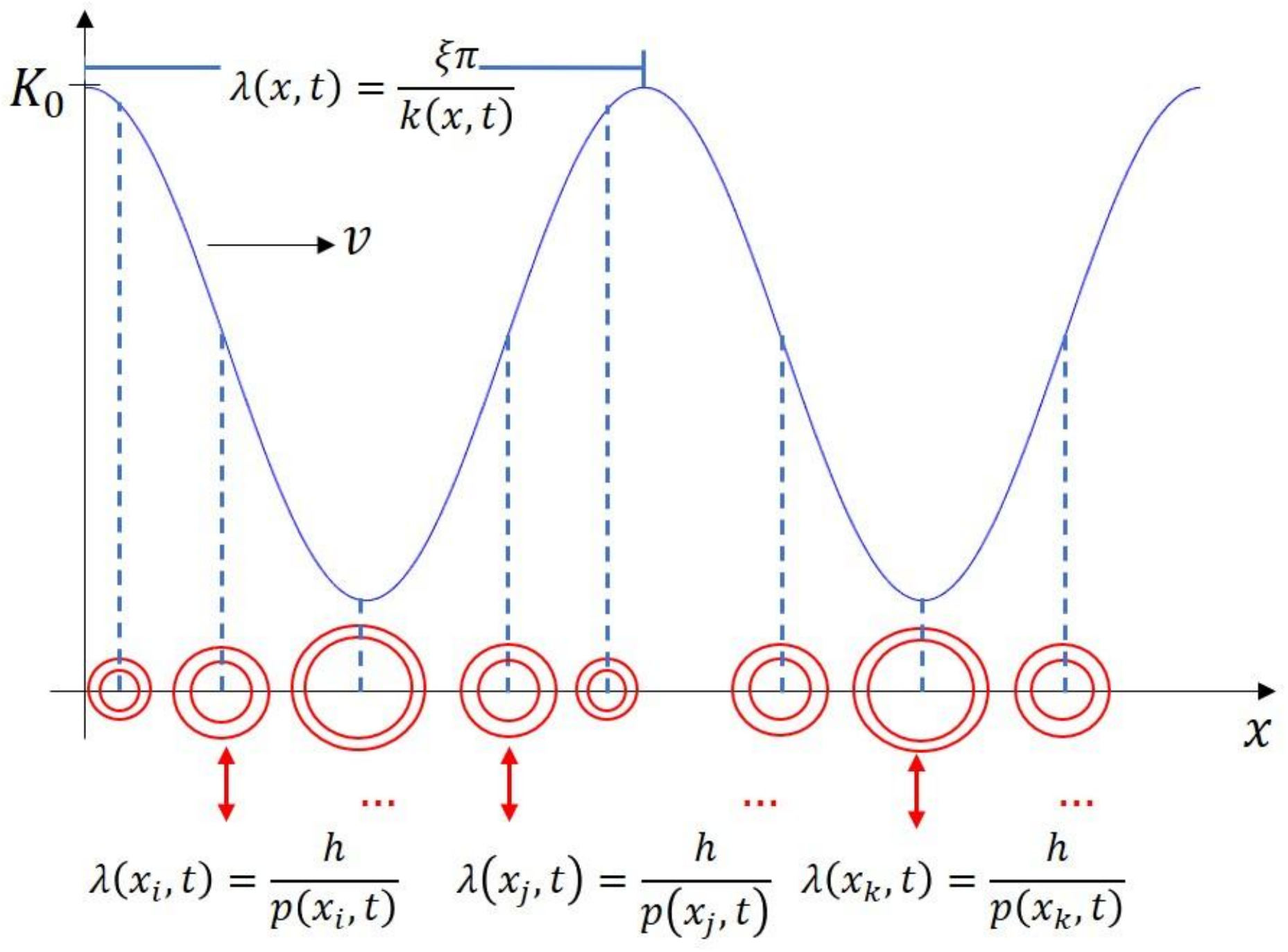

**FIG. 1.** The wavelength $\lambda = \frac{\xi\pi}{k}$ of a travelling wave with velocity $v$, originated by oscillating spherical membranes according to the de Broglie relation $\lambda = \frac{h}{p}$.

From such assumptions, it was derived an expression (Eq. (1)) with dimensions of energy

$$V_{ST} = \frac{hc}{\xi\pi x} \tag{1}$$

$V_{ST}$ could be well called potential, emerged due to the interaction between spherical membranes and a toroidal membrane $(ST)$ when $a \to 0$, $\omega_S = \omega_T$ while $k_S - k_T \to 0$. This mechanism was called compactification process. In this framework we call interaction particles those with $a \to 0$ and $\omega_S = \omega_T$. Equation (1) corresponds just now to the interaction between two physical elementary particles, in which it is assigned the dimensionless value 1 to the measurement unit of that would be an electric charge or mass. The value for any another greater particle being an integer multiple of that dimensionless value 1 given to the elementary particle (either electric charge or mass). The kind of potential (electric potential or gravitational potential) depends on the choice of the value for $\xi$ as it will be seen later.

Equation (1) can be derived from the Yukawa proposed free – space equation (Eq. (2))

$$\left(\nabla^2 - \frac{\partial^2}{c^2\partial t^2} - \frac{1}{a^2}\right)V(r,t) = 0 \tag{2}$$

which is certainly relativistically invariant and $a$ is a range parameter. Furthermore, Yukawa took the decisive step of treating $V$ quantum mechanically, by looking for a (de Broglie – type) propagating wave solution of Eq. (3), namely

$$V \propto \exp\left(\frac{i\,\vec{p}.\vec{r}}{\hbar} - \frac{iE\,t}{\hbar}\right) \tag{3}$$

For simplicity we write Eq. (3) in the one-dimensional case

$$V \propto \exp\left(\frac{i\,\vec{p}.\vec{r}}{\hbar} - \frac{iEt}{\hbar}\right) = \exp\left(\frac{i\,p\,x}{\hbar} - \frac{iE\,t}{\hbar}\right) = \exp\left(i\left(\frac{h}{\lambda}\frac{x}{\hbar} - \frac{\omega\hbar t}{\hbar}\right)\right)$$

$$= \exp\left(i\left(\frac{2\pi}{\lambda}\frac{h\,x}{h} - \frac{2\pi}{T}t\right)\right) = \exp\left(i\left(\frac{2\pi}{\lambda}x - \frac{2\pi}{T}t\right)\right) = \exp\left(2\pi i\left(\frac{x}{\lambda} - \frac{t}{T}\right)\right)$$

$$V \propto \exp\left(\frac{i\,\vec{p}\,.\,\vec{r}}{\hbar} - \frac{iEt}{\hbar}\right) = \exp\left(2\pi i\left(\frac{x}{\lambda} - \frac{t}{T}\right)\right)$$

At this point, we replace $\xi$ instead of 2, such that

$$V \propto \exp\left(\xi \pi i\left(\frac{x}{\lambda}-\frac{t}{T}\right)\right)$$

From this last equation we can write again

$$V \propto \exp\left(\frac{\xi}{2} i\left(\frac{2\pi}{\lambda} x-\frac{2\pi}{T} t\right)\right)=\exp\left(\frac{\xi}{2} i\left(\frac{2\pi x}{\frac{h}{p}}-\omega t\right)\right)=\exp\left(\frac{\xi}{2} i\left(\frac{px}{\hbar}-\frac{Et}{\hbar}\right)\right)$$

$$V \propto \exp\left(\frac{\xi}{2} i\left(\frac{px}{\hbar}-\frac{Et}{\hbar}\right)\right) \qquad (4)$$

Inserting Eq. (4) into Eq. (2) one finds

$$E=\sqrt{p^2c^2+\frac{4c^2\hbar^2}{\xi^2a^2}}$$

Comparing this with the standard $E-p$ relation for a massive particle in special relativity

$$E=\sqrt{p^2c^2+m^2c^4}$$

The quantum of the finite – range force field $V$ has a mass given by

$$m^2c^4=\frac{4c^2\hbar^2}{\xi^2a^2}$$

$$m=\frac{2\hbar}{\xi ca} \qquad (5)$$

When $\xi=2$ and $a=2\,fm$, we have $m\sim 100MeV$, which is the Yukawa's prediction for the mass of the nuclear force quantum.

From Eq. (5) we can write

$$V=mc^2=\frac{hc}{\xi\pi a}$$

and then,

$$V = \frac{hc}{\xi \pi x}$$

where the range parameter is replaced in general by $x$. This expression is the same given in Eq. (1). Now, as a consequence, we have a corresponding expression with force dimensions,

$$F_{ST} = \frac{hc}{\xi \pi x^2} \tag{6}$$

The relative intensity between Eq. (6) and the electrostatic force given by $F_C = \frac{K_e e^2}{x^2}$ is

$$\frac{F_{ST}}{F_C} = \frac{\frac{hc}{\xi \pi x^2}}{\frac{K_e e^2}{x^2}} = \frac{hc}{K_e \xi \pi e^2} = \frac{274.073487}{\xi}$$

$$\frac{F_{ST}}{F_C} = \frac{274.073487}{\xi} \tag{7}$$

Then, from Eq. (7) and since $\xi$ is an integer greater than or equal to 1, we have:

If $\xi = 1$, then $\frac{F_{ST}}{F_C} \sim 274$

If $\xi = 2$, then $\frac{F_{ST}}{F_C} \sim 137$ ($F_{ST}$ corresponds to the nuclear force)

If $\xi = 3$, then $\frac{F_{ST}}{F_C} \sim 91$

…

If $\xi = 274$, then $\frac{F_{ST}}{F_C} \sim 1$ ($F_{ST}$ corresponds to the electrostatic force)

…

If $\xi = 1.14198\times10^{45}$ ,then $\frac{F_{ST}}{F_C} \sim 2.3999719627595\times10^{-45}$

(In this case $F_{ST}$ corresponds to the gravitational interaction force)

Here we must emphasize the fact that when $\xi = 274$, the coulomb potential (repulsive energy) between two electrons given by $V = \frac{e^2}{4\pi\varepsilon_0 x}$ is equal to $V = \frac{hc}{274\pi x}$ according to Eq. (1). This means that

$$\frac{hc}{274\pi x} = \frac{e^2}{4\pi\varepsilon_0 x}$$

from which

$$\frac{hc}{2(137)\pi} = \frac{e^2}{4\pi\varepsilon_0}$$

$$\alpha\hbar c = \frac{e^2}{4\pi\varepsilon_0}$$

$$\alpha = \frac{e^2}{4\pi\varepsilon_0\hbar c}$$

This last expression is the fine structure constant. According to Zegarra et al.[8], the energy between two electrons would be 137 times the non-fundamental period $2\pi$ of the oscillation of membranes or 274 times the fundamental period, smaller than the energy of a photon with wavelength equal to the distance between two electrons. Moreover, 274 is a particular value that the parameter $\xi$ of unification of the formulas for the Coulomb's law and Newton's law of universal gravitation takes. In this context, the meaning of 137 turns out to be clear and hence the meaning of $\alpha$ . Figure 2, shows the graph of $K = K_0 \sin^2(kx - \omega t)$ with fundamental period $\pi$ given in second assumption and, non-fundamental period $2\pi$ such that 137 times $2\pi$ is the period of oscillation of the interacting membranes given in first assumption. Interacting membranes responsible for the interaction force between electrons.

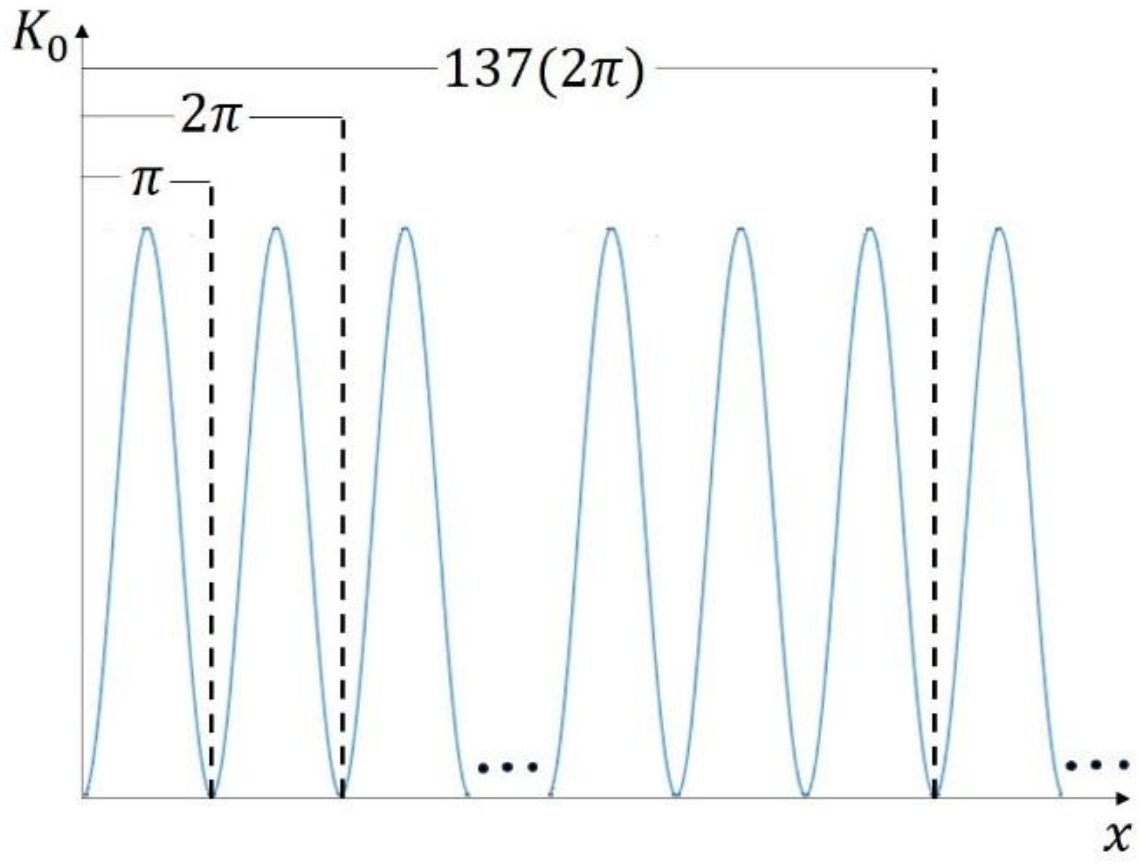

**FIG. 2.** Graph of $K = K_0 \sin^2(kx - \omega t)$ with fundamental period $\pi$ and non – fundamental period $2\pi$ such that 137 times $2\pi$ is the period of oscillation of interacting membranes which are responsible for the interaction force between electrons.

**2.1 Planck Mass**

Suppose that a mass $\Omega_\xi$ is composed of $n^{\Omega_\xi}$ elementary particles, where superscript refers to mass $\Omega_\xi$, and that the interaction force $F_{ST}$ for two identical masses is $F_{ST} = \dfrac{hc\left(n^{\Omega_\xi}\right)^2}{\xi \pi x^2}$ according to the formula deduced from the torus – sphere interaction. On the other hand, suppose that $F_G = \dfrac{Gm^2}{x^2}$ corresponds to the interaction between two bodies of equal mass $m$ under the Newton's law of universal gravitation. Suppose the relative intensity between the two forces above equals 1 for any mass $m$. Then, with the purpose of replacing the classic formula of gravitational interaction $F_G$ given by Newton, by the new $F_{ST}$, we must have

$$\frac{F_{ST}}{F_G} = \frac{\dfrac{hc\left(n^{\Omega_\xi}\right)^2}{\xi\pi x^2}}{\dfrac{Gm^2}{x^2}} = 1$$

from which,

$$\frac{hc\left(n^{\Omega_\xi}\right)^2}{\xi\pi x^2} = \frac{Gm^2}{x^2}$$

and hence,

$$n^{\Omega_\xi} = \sqrt{\frac{G\xi\pi}{hc}}m$$

Since $n^{\Omega_\xi}$ is the number of elementary particles present in each interacting body of mass $\Omega_\xi$, we demand to determine a certain elementary mass $M_e$ in $m$, in such a way $m = n^{\Omega_\xi} M_e$. Then, we have that

$$M_e = \frac{m}{n^{\Omega_\xi}} = \frac{m}{\sqrt{\frac{G\xi\pi}{hc}}m}$$

$$M_e = \frac{m}{n^{\Omega_\xi}} = \frac{m}{\sqrt{\frac{G\xi\pi}{hc}}m}$$

$$M_e = \sqrt{\frac{hc}{\xi\pi G}}$$

Given values to $\xi$:

If $\xi = 1$, then $M_e = \sqrt{\frac{hc}{\pi G}}$

If $\xi = 2$, then $M_e = \sqrt{\dfrac{hc}{2\pi G}}$ (8)

Equation (8) corresponds to the Planck's mass $M_P$ as it is well known in ordinary literature.

Say,

$$M_P = M_e = \sqrt{\frac{hc}{2\pi G}} = 2.176434 \times 10^{-8} kg$$ [1]

This Value is in accordance with the value deduced from dimensional analysis.

On the other hand, Zegarra, et al.[8] deduced the following equation

$$\frac{\sin^2\left(k_T x - \omega_T t\right)}{\sin^2\left(k_s x - \omega_s t\right)} = 1 \tag{9}$$

and hence,

$$k_S = k_T \ , \ \omega_S = \omega_T \tag{10}$$

when an annihilation process has happened.

In order to determine the way in which $\left(\omega_S - \omega_T\right) \to 0$, along the temporal axis and satisfy Eq. (10) when $k_S = k_T = k$, Eq. (9) implies

$$\sin\left(kx - \omega_T t\right) = \sin\left(kx - \omega_S t + \sigma\right) \tag{11}$$

With $\sigma \neq 0$. Then, Eq. (11) implies

$$kx - \omega_T t = kx - \omega_S t + \sigma$$

from which

$$\omega_S - \omega_T = \frac{\sigma}{t} \tag{12}$$

[1] According to CODATA 2018 https://physics.nist.gov, the value is $2.176434 \times 10^{-8} kg$

It is well known that in a uniform circular movement, $\omega = \frac{2\pi}{T}$, where $2\pi$ is the minimum complete revolution. According to Zegarra, et al.[8], for a minimum complete revolution (fundamental), $\omega_F = \frac{\pi}{T}$ and in general $\omega = \frac{\zeta\,\pi}{T}$ for any number of complete revolutions, being $\zeta$ an integer. Then,

$$\omega_S - \omega_T = \frac{\zeta_S \pi}{T_S} - \frac{\zeta_T \pi}{T_T} = \frac{\sigma}{t} \quad (13)$$

If $\zeta = \zeta_S = \zeta_T$, Eq. (13) becomes

$$\left(\frac{\pi}{T_S} - \frac{\pi}{T_T}\right) = \frac{\sigma}{\zeta t} \quad (14)$$

If $\omega_{S_F} = \frac{\pi}{T_S}$ , $\omega_{T_F} = \frac{\pi}{T_T}$

Eq. (14) is written as

$$\omega_{S_F} - \omega_{T_F} = \frac{\sigma}{\zeta t} \quad (15)$$

If $\omega^* = \omega_{S_F} - \omega_{T_F}$

through Eq. (15), Eq. (14) is reduced to

$$\omega^* = \frac{\sigma}{\zeta t} \quad (16)$$

According to Zegarra et al.[8], if $\sigma = 1$ in Eq. (16), we have,

$$\omega^* = \frac{1}{\zeta t} \quad (17)$$

from which,

$$t = \frac{1}{\omega^* \zeta} \quad (18)$$

Furthermore, from $V=\dfrac{hc}{\zeta\pi x}$ (Eq. (1)) we have

$$x=\frac{hc}{\xi\pi V} \tag{19}$$

Fifth assumption (additional to the previous four)**:** The relation $x=ct$ holds, with $c$ the light velocity.

According to Eq. (18) and Eq. (19) and the fifth assumption, we have that

$$\frac{x}{t}=\frac{\dfrac{hc}{\zeta\pi V}}{\dfrac{1}{\omega^{*}\zeta}}=\frac{hc\zeta\omega^{*}}{\zeta\pi V}=c$$

Say,

$$\frac{h\omega^{*}}{\pi V}=1 \tag{20}$$

From Eq. (21)

$$\omega^{*}=\frac{\pi V}{h} \tag{21}$$

From Eqs. (17) and (21)

$$\frac{\pi V}{h}=\frac{1}{\zeta t}$$

Hence,

$$t=\frac{h}{\zeta\pi V} \tag{22}$$

**2.2 The Planck's time**

The value of the Planck's time can be obtained from two different ways:

**(a)** By using Eq. (22) and taking the energy of the Planck mass equal to $1.220890\times10^{19}\pm0.000014\times10^{19}\,GeV$, equivalent in Joules to $1.956081\times10^{19}\,J$, as the value for the potential $V$, in addition to $\zeta=2$ and $h=6.62607015\times10^{-34}\,J\,s$. This let us obtain from Eq. (22) that the Plank's time $t_P$ is,

$$t_P=\frac{6.62607015\times10^{-34}\,J\,s}{2\times\pi\times1.956081\times10^{9}\,J}=5.391248\times10^{-44}\,s$$

**(b)** By using the Heisenberg uncertainty relation, it is well known that $\Delta E.\Delta t=\hbar$. By rewritten this last relation as $E_P.t_P=\hbar$ we have $t_P$ such that,

$$t_P=\frac{h}{2\pi E_P}=\sqrt{\frac{hG}{2\pi c^5}}=5.391248\times10^{-44}\,s$$

where

$$E_P=M_Pc^2=\sqrt{\frac{hc^5}{2\pi G}}$$

**2.3 The Planck's length**

By using the fifth assumption and $c=299792458\,m\,s^{-1}$[2] together with the value of the Planck's time given above, we have that the Planck's length $L_P$ is,

$$L_P=ct=\left(2.99792458\times10^{8}\,\frac{m}{s}\right)\left(5.391248\times10^{-44}\,s\right)=16.162555\times10^{-36}\,m$$

$$L_P=16.162555\times10^{-36}\,m$$

$$L_P=1.6162555\times10^{-33}\,cm$$

[2] CODATA: https://physics.nist.gov

which is also the same as

$$L_P = ct_P = \frac{ch}{2\pi E_P} = \sqrt{\frac{hG}{2\pi c^3}} = 1.6162555\times 10^{-33}\,cm$$

**3. Discussion**

Here, the Planck's mass is obtained as the result of equating two formulas:

$F_{ST} = \frac{hc\left(n^{\Omega_\xi}\right)^2}{\xi\pi x^2}$, which corresponds to the interaction between the torus and sphere membranes considered, where $n^{\Omega_\xi}$ is the number of elementary particles present in any mass $\Omega_\xi$ and $\xi$ is a dimensionless parameter such that $\frac{\xi\pi}{k}$ is the spatial period with which oscillate the curvatures of such membranes[8]. And the other formula, the Newton's law of universal gravitation given by $F_G = \frac{Gm^2}{x^2}$. It was found that the elementary mass $M_e$ such that $m = n^{\Omega_\xi} M_e$, results to be $\sqrt{\frac{hc}{2\pi G}}$, which is the Planck's mass when $\xi = 2$.

The Planck's time is obtained when $k_S = k_T$ and it is determined from the way in which $(\omega_S - \omega_T) \to 0$ along the time axis. It was obtained $t = \frac{h}{\xi\pi V}$ where $\zeta$ is a dimensionless parameter such that $\omega = \frac{\zeta\pi}{T}$ is the angular frequency with which oscillate the corresponding curvatures of the torus and sphere membranes. With the help of $E_P = M_P c^2 = \sqrt{\frac{hc^5}{2\pi G}}$, $\zeta = 2$ and $t = \frac{h}{\zeta\pi V}$ it is obtained $t_P = \frac{h}{2\pi E_P} = \sqrt{\frac{hG}{2\pi c^5}}$, with $V = E_P$.

The Planck's length $L_P$, results of assuming that for relativistic particles we have $\frac{x}{t}=c$. Say, $L_P=ct_P$. This means that both $t_P$ and $L_P$ take place when $\xi=2$ and $\zeta=2$. The Planck's mass takes place when $\xi=2$ regardless of how much $\zeta$ is worth. See figure 3.

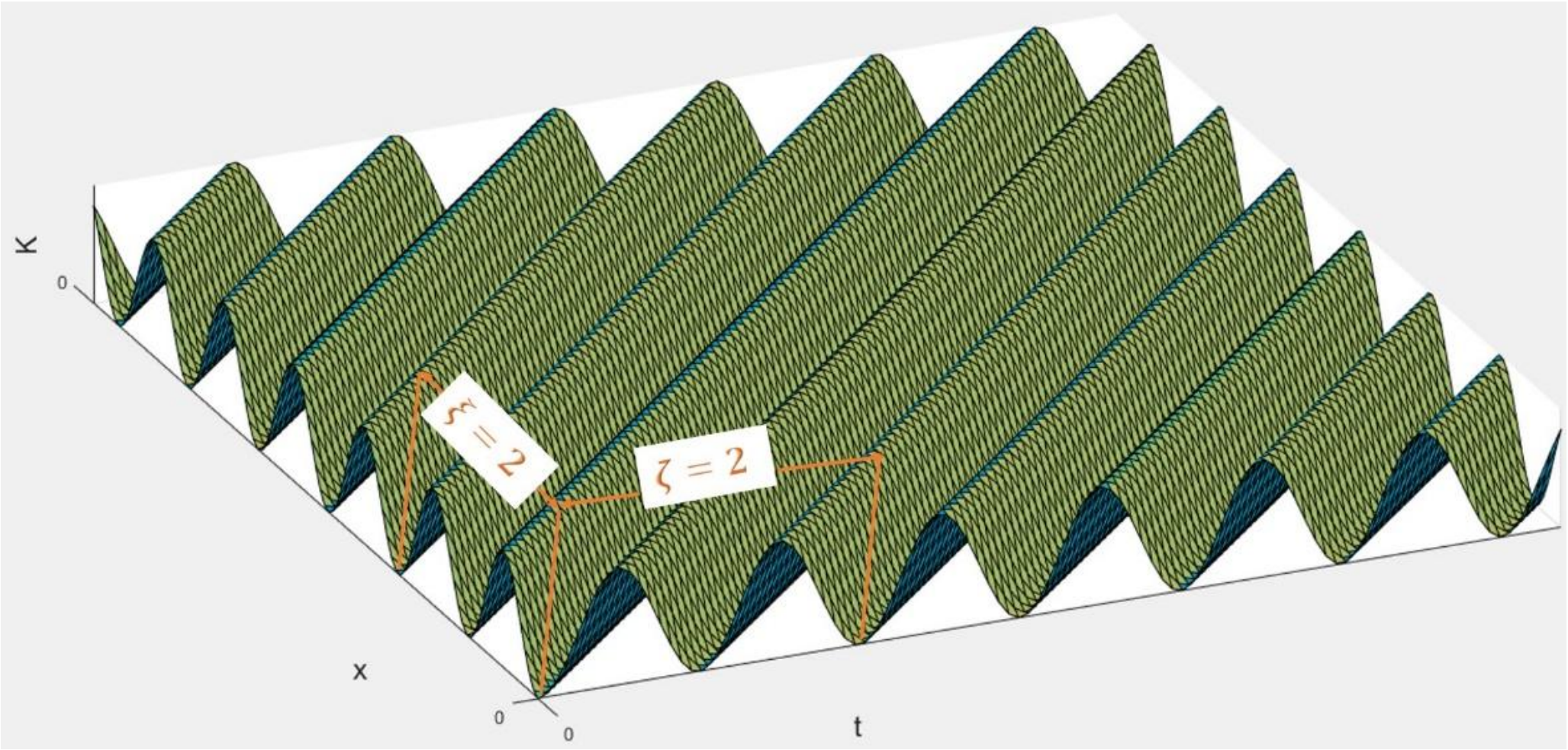

**FIG. 3.** Surface representing $K=K_0\sin^2\left(kx-\omega t\right)$. Both $t_P$ and $L_P$ take place when $\xi=2$ and $\zeta=2$. The Planck's mass takes place when $\xi=2$ regardless of how much $\zeta$ is worth. The periods of oscillation are close to the fundamental period $\pi$ along both the $x-$axis and the $t-$axis.

The expressions for the Planck's units obtained here are consistent with the current ones in which the number $2\pi$ appears under the denominator of the expressions below the radicals. Say, now the quantum for the angular momentum of quantum mechanics $\hbar=\frac{h}{2\pi}$ is present in the expressions of the Planck's units from a natural way. Originally, Planck[10] wrote his units as $M_P = \sqrt{\frac{bc}{f}}$, $t_P = \sqrt{\frac{bf}{c^5}}$ y $L_P = \sqrt{\frac{bf}{c^3}}$ or in a modern notation as

$$M_P = \sqrt{\frac{hc}{G}},\ t_P = \sqrt{\frac{hG}{c^5}},\ L_P = \sqrt{\frac{hG}{c^3}},\ (b = h,\ f = G)$$

A discussion of the various ordinary methods of determining the expressions for the Planck units such as mass, length and time, can be found in Crothers and Dunning-Davies.[3] They mention three methods: dimensional analysis, equating Compton wavelength and Schwarzschild radius and, the quantum/gravity approach. They state that the first method neither contains no physics nor does it try to contain it. In the case of the second and third methods, on the other hand, do seem to contain some physics, but via a closer examination they doubt about that. Furthermore, they emphasize the lack of the number $2\pi$ when deriving Planck mass via dimensional analysis and, the lack of only the number $\pi$ in deriving Planck mass via the equality of Compton wavelength and Schwarzschild radius. When using membranes model, the presence of $2\pi$ is a consequence of supposing that the curvature of the membranes is oscillating with spatial period $\frac{\xi\pi}{k}$, when $\xi = 2$, value associated with a sort of nuclear force. This nuclear force, is given by Eq. (6) when $\xi = 2$.

## 4. Conclusions

Within the framework of the theory of Zegarra, et al.,[8] based on the concept of membrane fundamentally, the values found for the Planck's mass, Planck's time and Planck's length, were respectively

$$M_P = \sqrt{\frac{hc}{2\pi G}} = 2.176434\times 10^{-8}\,kg$$

$$t_P = \frac{h}{2\pi E_P} = \sqrt{\frac{hG}{2\pi c^5}} = 5.391248\times10^{-44}\, s$$

$$L_P = ct_P = \frac{ch}{2\pi E_P} = \sqrt{\frac{hG}{2\pi c^3}} = 1.6162555\times10^{-33}\, cm$$

where $E_P = M_P c^2 = \sqrt{\frac{hc^5}{2\pi G}}$

These values are in accordance with those already known. These expressions are related to the parameter $\xi$ that determine the period of oscillation of the membranes given in the first assumption above, and it is a parameter of unification between the formulas for the Coulomb's law and Newton's law of universal gravitation linking the forces of microworld and macroworld. Depending on the value $\xi$ takes, one force or another will be had. Since in this case, the value for the parameter is 2, closely to minimum value 1, the Planck units arise there where the membranes oscillate with period $2\pi$ closely to the fundamental period $\pi$. In this context, Planck units must be fundamental. For $\xi = 2$ and when the membranes given above interact, causes a force like as the nuclear force.

**Acknowledgments**

The authors would like to thank to *Universidad Nacional de Chimbote – UNS,* Chimbote-Perú. Universidade Federal do Pampa – Unipampa and to the Brazilian agency: Conselho Nacional de Desenvolvimento Científico e Tecnologico (CNPq) through the scholarships in Research Productivity of Luis E. G. Armas.